\pdfoutput=1
\documentclass{comex}
\usepackage[T1]{fontenc}
\usepackage[pass]{geometry}
\usepackage{stix}
\usepackage[a-2u]{pdfx}

\newtheorem{theorem}{Theorem}
\newtheorem{definition}[theorem]{Definition}
\newtheorem{remark}[theorem]{Remark}

\title{Rate-Distortion-Perception Tradeoff of Variable-Length Source Coding for General Information Sources\thanks{%
    The result in this letter will be orally presented at
    poster session of
the 41st Symposium on Information Theory and its Applications,
Fukushima, Japan, 18--21 December 2018.}}
\author{Ryutaroh Matsumoto$^{\rm 1a,2)}$}
\affiliate{$^{1}$
  Department of Information and Communication Engineering,
  Nagoya University\\
Furo-cho, Chitane-ku, Nagoya, Aichi 464-8603, Japan.\\
$^{2}$
  Department of Mathematical Sciences, Aalborg University, Denmark.}

\email{{\rm a)} ryutaroh.matsumoto@nagoya-u.jp}
\received{2018}{11}{3}
\accepted{2018}{11}{30}
\vol{8}
\no{3}

\begin{document}
\sloppy
\vol{8}
\no{3}
\maketitle

\begin{abstract}
Blau and Michaeli recently introduced a novel concept
for inverse problems of signal processing, that is,
the perception-distortion tradeoff.
We introduce their tradeoff into the rate distortion theory
of variable-length lossy source coding in information theory,
and clarify the tradeoff among information rate, distortion
and perception for general information sources.
We also discuss the fixed-length coding with average distortion
criterion that was missing in the previous letter.
\end{abstract}
\begin{keywords}
  perception-distortion tradeoff, rate-distortion theory, data compression,
  variable-length coding
\end{keywords}
\begin{classification}
Fundamental theories for communications
\end{classification}


\urlstyle{tt}

\section{Introduction}
An inverse problem of signal processing
is to reconstruct the original information from its degraded
version. It is not limited to image processing, but it often
arises in the image processing.
When a natural image is reconstructed, the reconstructed
image sometimes does not look natural while it is close to
the original image by a reasonable metric, for example
mean squared error.
When the reconstructed information is close to the original,
it is often believed that it should also look natural.

Blau and Michaeli \cite{cvpr2018}
questioned this unproven belief.
In their research \cite{cvpr2018},
they mathematically formulated the \emph{naturalness}
of the reconstructed information by a
distance of the probability distributions of
the reconstructed information and the original information.
The reasoning behind this is that the perceptional quality of
a reconstruction method is often evaluated by
how often a human observer can distinguish an output of
the reconstruction method from natural ones.
Such a subjective evaluation can mathematically be
modeled as a hypothesis testing \cite{cvpr2018}.
A reconstructed image is more easily distinguished
as the variational distance
$\sigma(P_R$, $P_N)$ increases \cite{cvpr2018},
where $P_R$ is the probability distribution of
the reconstructed information and
$P_N$ is that  of
the natural one.
They regarded the perceptional quality of reconstruction as
a distance between $P_R$ and $P_N$.
The distance between the reconstructed information
and the original information is conventionally called as  distortion.
They discovered that there exists a tradeoff
between perceptional quality and distortion, and
named it as the \emph{perception-distortion tradeoff}.

Claude Shannon \cite[Chapter 5]{hanbook} initiated the rate-distortion
theory in 1950's.
It clarifies the tradeoff between
information rate and distortion in the lossy source coding
(lossy data compression).
The rate-distortion theory has served as a theoretical
foundation of image coding for past several decades, as
drawing a rate-distortion curve is a common practice in
research articles of image coding.
Since distortion and perceptional quality are now considered
two different things,
it is natural to consider a tradeoff among
information rate, distortion and perceptional quality.
Blau and Michaeli \cite{cvpr2018}
briefly mentioned the rate-distortion theory,
but they did not clarify the tradeoff among the three.
Then the author \cite{matsumotocomex} clarified the tradeoff
among the three for fixed-length coding, but did not
clarified variable-length coding, where
fixed and variable refer the length of a codeword is
fixed or variable, respectively \cite[Chapter 5]{hanbook}.
The variable-length lossy source coding is practically
more important than
the fixed-length counterpart because
most of image and audio coding methods are variable-length.

The purpose of this letter is to mathematically
define the tradeoff of variable-length lossy source
coding for general information sources,
and to express the tradeoff in terms of
information spectral quantities introduced by Han and Verd\'u
\cite{hanbook}.
We also discuss the fixed-length coding with average distortion
criterion that was missing in the previous letter \cite{matsumotocomex}.

Since the length limitation is strict in this journal,
citations to the original papers are replaced by
those to the textbook \cite{hanbook}, and
the mathematical proof is a bit compressed.
The author begs readers' kind understanding.
The base of $\log$ is an arbitrarily fixed real number $>1$
unless otherwise stated.

\section{Preliminaries}
The following definitions are borrowed from Han's textbook
\cite{hanbook}.
Let
\[
\mathbf{X} = \left\{ X^n = (X_1^{(n)}, \ldots, X_n^{(n)} ) \right\}_{n=1}^\infty
\]
be a general information source, where the alphabet of the random variable $X^n$
is the $n$-th Cartesian product $\mathcal{X}^n$ of some finite alphabet
$\mathcal{X}$.
For a sequence of real-valued random variables $Z_1$, $Z_2$, \ldots
we define
\[
\textrm{p-}\limsup_{n\rightarrow \infty} Z_n =\inf\left\{
\alpha \mid \lim_{n\rightarrow\infty} \mathrm{Pr}[Z_n > \alpha] = 0 \right\}.
\]
For two general information sources $\mathbf{X}$ and $\mathbf{Y}$ we define
\begin{eqnarray*}
\overline{I}(\mathbf{X}; \mathbf{Y}) &=& \textrm{p-}\limsup_{n\rightarrow \infty}
\frac{1}{n} \log \frac{P_{X^nY^n}(X^n, Y^n)}{P_{X^n}(X^n)P_{Y^n}(Y^n)},\\
H_K(\mathbf{X}) &=& \limsup_{n\rightarrow \infty}
\frac{1}{n} H_K(X^n),\\
F_\mathbf{X}(R)& =& \limsup_{n\rightarrow \infty} \mathrm{Pr}\left[
  \frac{1}{n} \log \frac{1}{P_{X^n}(X^n)} \geq R \right],
\end{eqnarray*}
where $H_K(X^n)$ is the Shannon entropy of $X^n$ in $\log_K$.

For two distributions $P$ and $Q$ on an alphabet $\mathcal{X}$,
we define the variational distance $\sigma(P,Q)$ as
$\sum_{x\in \mathcal{X}} |P(x)-Q(x)|/2$. In the rate-distortion theory, we usually
assume a reconstruction alphabet different from a source alphabet.
In order to consider the distribution similarity of reconstruction,
in this letter we assume $\mathcal{X}^n$ as both source and
reconstruction alphabets.

\section{Variable-length source coding}\label{sec2}
An encoder of length $n$ is a \emph{stochastic}
mapping $f_n: \mathcal{X}^n \rightarrow
\mathcal{U}^*$, where $\mathcal{U} = \{1$, \ldots, $K\}$
and $\mathcal{U}^*$ is the set of finite-length sequences over $\mathcal{U}$.
By \emph{stochastic} we mean that the encoder output $f_n(x^n)$
is probabilistic with a fixed input $x^n \in \mathcal{X}^n$.
The corresponding decoder of length $n$
is a deterministic mapping
$g_n: \mathcal{U}^* \rightarrow \mathcal{X}^n$.
We denote by $|f_n(x^n)|$ the (random variable of)
length of sequence $f_n(x^n) \in \mathcal{U}^*$
for $x^n \in \mathcal{X}^n$.
We denote by $\delta_n: \mathcal{X}^n \times \mathcal{X}^n
\rightarrow [0,\infty)$ a general distortion function.

  \subsection{Average distortion criterion}
\begin{definition}
A triple $(R,D,S)$ is said to be $va$-achievable
if there exists a sequence of encoder and decoder
$(f_n$, $g_n)$ such that
\begin{eqnarray}
  \limsup_{n\rightarrow \infty} \frac{\mathbf{E}[\log |f_n(X^n)|]}{n} & \leq & R,\label{eq3}\\
  \limsup_{n\rightarrow \infty} \frac{1}{n}\mathbf{E}[\delta_n(X^n, g_n(f_n(X^n)))] & \leq & D,\label{eq4}\\
  \limsup_{n\rightarrow \infty} \sigma(P_{g_n(f_n(X^n))}, P_{X^n}) & \leq & S.\label{eq5}
\end{eqnarray}
Define the function $R_{va}(D,S)$ by
\[
R_{va}(D,S) = \inf\{ R \mid (R,D,S) \mbox{ is $va$-achievable }\}.
\]
\end{definition}

\begin{theorem}\label{thm1}
  \[
  R_{va}(D,S) = \inf_{\mathbf{Y}} H_K(\mathbf{Y})
  \]
  where the infimum is taken with respect to all
  general information sources $\mathbf{Y}$ satisfying
\begin{eqnarray}
  \limsup_{n\rightarrow \infty} \frac{1}{n}\mathbf{E}[\delta_n(X^n, Y^n)] & \leq & D,\label{eq6}\\
  \limsup_{n\rightarrow \infty} \sigma(P_{Y^n}, P_{X^n}) & \leq & S.\label{eq7}
\end{eqnarray}
\end{theorem}

\noindent\textbf{Proof:}
Let a pair of encoder $f_n$ and decoder $g_n$ satisfies
Eqs.\ (\ref{eq3})--(\ref{eq5}).
Let $Y^n=g_n(f_n(X^n))$, and define the general information
source $\mathbf{Y}$ from $Y^n$.
We immediately see that $\mathbf{Y}$ satisfies Eqs.\ (\ref{eq6})
and $(\ref{eq7})$.
By the same argument as
\cite[p.\ 349]{hanbook}
we immediately see
\[
\limsup_{n\rightarrow \infty} \frac{\mathbf{E}[\log |f_n(X^n)|]}{n} \geq H_K(\mathbf{Y}).
\]

On the other hand,
suppose that a general information source
$\mathbf{Y}$ satisfies Eqs.\ (\ref{eq6}) and (\ref{eq7}).
Let $f'_n$ and $g'_n$ be a \emph{lossless} variable-length
encoder and its  decoder \cite[Section 1.7]{hanbook}
for $\mathbf{Y}$
such that $Y^n = g'_n(f'_n(Y^n))$ and 
\[
\limsup_{n\rightarrow \infty} \frac{\mathbf{E}[\log |f'_n(Y^n)|]}{n} = H_K(\mathbf{Y}).
\]
For a given information sequence $x^n \in \mathcal{X}^n$,
the encoder randomly chooses $y^n\in \mathcal{Y}^n$ according to the
conditional distribution $P_{Y^n|X^n}(\cdot | x^n)$,
and define the codeword as $f'_n(y^n)$.
The decoding result is $y^n=g'_n(f'_n(y^n))$.
Since the probability distribution of
decoding result $g'_n(f'_n(y^n))$ is $P_{Y^n}$,
we see that the constructed encoder and decoder
satisfy Eqs.\ (\ref{eq4}) and (\ref{eq5}).
\rule{1ex}{1ex}

\subsection{Maximum distortion criterion}
\begin{definition}
A triple $(R,D,S)$ is said to be $vm$-achievable
if there exists a sequence of encoder and decoder
$(f_n$, $g_n)$ such that
\begin{eqnarray*}
  \limsup_{n\rightarrow \infty} \frac{\mathbf{E}[\log |f_n(X^n)|]}{n} & \leq & R,\\
  \textrm{p-}\limsup_{n\rightarrow \infty} \frac{1}{n}\delta_n(X^n, g_n(f_n(X^n))) & \leq & D,\\
  \limsup_{n\rightarrow \infty} \sigma(P_{g_n(f_n(X^n))}, P_{X^n}) & \leq & S.
\end{eqnarray*}
Define the function $R_{vm}(D,S)$ by
\[
R_{vm}(D,S) = \inf\{ R \mid (R,D,S) \mbox{ is $vm$-achievable }\}.
\]
\end{definition}

\begin{theorem}\label{thm2}
  \[
  R_{vm}(D,S) =  \inf_{\mathbf{Y}} H_K(\mathbf{Y})
  \]
  where the infimum is taken with respect to all
  general information sources $\mathbf{Y}$ satisfying Eq.\ (\ref{eq7}) and
\[
  \textrm{p-}\limsup_{n\rightarrow \infty} \frac{1}{n}\delta_n(X^n, Y^n)  \leq  D.
\]
\end{theorem}

\noindent\textbf{Proof:}
The proof is almost the verbatim copy of that of Theorem \ref{thm1} and
is omitted.
\rule{1ex}{1ex}

\begin{remark}
  The tradeoff for variable-length coding
  with the average distortion criterion and without
  the perception criterion was also determined by
  using \emph{stochastic} encoders \cite[Section 5.7]{hanbook},
  but with the maximum distortion criterion without the perception
  criterion, only the \emph{deterministic} encoders were sufficient
  to clarify the tradeoff \cite[Section 5.6]{hanbook}.
  It is not clear at present whether or not
  we can remove the randomness from encoders in Theorem \ref{thm2}.
\end{remark}

\section{Fixed-length coding with the average distortion criterion}
In this section we state the tradeoff for
fixed-length coding with the average distortion criterion,
because it has never been stated elsewhere. The proof is
almost the same as \cite{matsumotocomex}.
Note that the definition of encoder will be
different from Section \ref{sec2} and that an assumption on
the distortion $\delta_n$ will be added.

An encoder of length $n$ is a \emph{deterministic}
mapping $f_n: \mathcal{X}^n \rightarrow
\{1$, \ldots, $M_n\}$, and
the corresponding decoder of length $n$
is a deterministic mapping
$g_n: \{1$, \ldots, $M_n\} \rightarrow \mathcal{X}^n$.
We require an additional assumption that
$\delta_n(x^n, x^n)=0$ for all $n$ and $x^n \in \mathcal{X}^n$.

\begin{definition}
A triple $(R,D,S)$ is said to be $fa$-achievable
if there exists a sequence of encoder and decoder
$(f_n$, $g_n)$ such that
\begin{eqnarray*}
  \limsup_{n\rightarrow \infty} \frac{\log M_n}{n} & \leq & R,\\
  \limsup_{n\rightarrow \infty} \frac{1}{n}\mathbf{E}[\delta_n(X^n, g_n(f_n(X^n)))] & \leq & D,\\
  \limsup_{n\rightarrow \infty} \sigma(P_{g_n(f_n(X^n))}, P_{X^n}) & \leq & S.
\end{eqnarray*}
Define the function $R_{fa}(D,S)$ by
\[
R_{fa}(D,S) = \inf\{ R \mid (R,D,S) \mbox{ is $fa$-achievable }\}.
\]
\end{definition}

\begin{theorem}
  \[
  R_{fa}(D,S) = \max\left\{ \inf_{\mathbf{Y}} \overline{I}(\mathbf{X}; \mathbf{Y}),
  \inf\{R \mid F_{\mathbf{X}}(R) \leq S\} \right\}
  \]
  where the infimum is taken with respect to all
  general information sources $\mathbf{Y}$ satisfying
  \[
    \limsup_{n\rightarrow \infty} \frac{1}{n}\mathbf{E}[\delta_n(X^n, Y^n)]  \leq  D.\label{eq1}
  \]
\end{theorem}

\noindent\textbf{Proof:}
Proof is almost the verbatim copy of that of \cite{matsumotocomex}.

\end{document}